\documentclass[pra,twocolumn,a4paper,showpacs,superscriptaddress]{revtex4}
\usepackage{amsmath}
\usepackage{amsfonts}
\usepackage{graphicx}
\usepackage{longtable}
\newcommand{\be}{\begin{equation}}
\newcommand{\ee}{\end{equation}}

\newcommand{\ba}[1]{\left(\begin{array}{#1}}
\newcommand{\ea}{\end{array}\right)}
\begin{document}

\title{Entropic characterization of separability in Gaussian states} 

\author{Sudha}
\email{arss@rediffmail.com}
\affiliation{Department of Physics, Kuvempu University, 
Shankaraghatta, Shimoga-577 451, India}
\author{A. R. Usha Devi}
\affiliation{Department of Physics, Bangalore University, 
Bangalore-560 056, India}
\affiliation{Inspire Institute Inc., McLean, VA 22101, USA.}
\author{A. K. Rajagopal} 
\affiliation{Inspire Institute Inc., McLean, VA 22101, USA.}
\date{\today}

\begin{abstract} 
We explore separability of bipartite divisions of mixed Gaussian states based on the positivity of the 
Abe-Rajagopal (AR) $q$-conditional entropy. The AR $q$-conditional entropic characterization provide more 
stringent restrictions on separability (in the limit $q\rightarrow \infty$) than that obtained from the 
corresponding von Neumann conditional entropy ($q=1$ case) --  similar to the situation in finite dimensional 
states. Effectiveness of this approach, in relation to the results obtained by partial transpose criterion, is 
explicitly analyzed in three illustrative examples of two-mode Gaussian states of physical significance.   
\end{abstract}
\maketitle

Characterizing separability of a multipartite quantum state is a central issue in the subject of quantum 
information. Given  density matrix of a composite system, it is hard to decide its separability status, solely 
based on its intrinsic properties. 
In 1989 Werner~\cite{RFW1}  defined {\em inseparability}  by pointing out 
the impossibility of expressing an entangled composite quantum state as a convex mixture of its subsystem 
states.  Peres~\cite{Peres} enunciated    
{\em positivity under partial transpose} (PPT) criterion for 
separability of  bipartite states based on this definition in 1996. The PPT  
criterion was soon shown to be both necessary and sufficient in finite dimensional $2\times 2$ and $2\times 3$ 
systems  by R. Horodecki~\cite{Hor}. Peres' criterion has also led to an often-used  quantifying measure  of 
entanglement viz., negativity/logarithmic negativity~\cite{RFW2}. 
Much of the work that followed ever since has been focused on identifying less formidable sufficient -- though 
not necessary -- conditions for separability, as well as other mathematical methods for 
their analysis, such as positive and completely positive maps. A comprehensive review of these works 
on finite dimensional discrete systems and less extensively on the continuous systems may be found in the recent 
review article by Horodecki et. al.~\cite{Hor2}. 

Besides finite dimensional discrete systems, the issue of separability in continuous 
variable (CV) composite states, such as coupled bosonic oscillator systems (light modes), belonging to  infinite 
dimensional Hilbert spaces, too has invited much attention. 
The importance of investigating CV systems is evidenced by the tremendous 
activity in this field, as is clear from the review articles on this topic~\cite{AKP}. Fortunately,  
Peres' criterion has been extended to bipartite CV states and is found to be  both necessary and sufficient 
for two-mode Gaussian states~\cite{RS,Duan}. 
In fact, Gaussian states form a distinguished class amongst the CV systems due to   
experimental and theoretical ease they offer. Logarithmic negativity~\cite{RFW2} has also been employed to 
quantify entanglement in multimode Gaussian states and it provides  a  necessary and sufficient way of 
characterizing entanglement in the case of two mode Gaussian states. Quantification of entanglement of 
two-mode Gaussian states  in terms of minimal set of local measurements and classical communication has 
been developed in Ref.~\cite{Rigolin}. Further, entanglement of formation  has been analytically 
computed for arbitrary two mode Gaussian states~\cite{Pau}.    

A physically elegant method to characterize separability is based on the use of global and local 
spectra of the composite quantum system -- which forms the basis of entropic approach for 
separability~(\cite{Cerf}-- \cite{vwolf}).  
Whereas the non-negativity of von-Neumann conditional entropy is used to identify entangled pure states, it is 
inadequate to address the issue of separability in
mixed states. Generalized entropic measures~(\cite{Cerf} --\cite{vwolf}) offer more 
sophisticated tools to explore global vs local disorder in mixed states and lead to  stringent limitation on 
separability than that obtained using positivity of von Neumann conditional entropy. 
Horodecki et. al.~\cite{HorEn} recognized that conditional Renyi 
entropies  are necessarily non-negative for all separable states, while they can assume negative values 
by  entangled states. Employing Tsallis entropy~\cite{tsallis}, indexed by a real parameter
$q\in[0,\infty]$, Abe and Rajagopal~\cite{AR} defined $q$-conditional entropy associated with the bipartite 
division of a density matrix $\rho(A,B)$ and its subsystem $\rho(A)={\rm Tr}_{B}\,[\rho(A,B)]$ as 
\begin{eqnarray}
\label{sba}
S_{q}(B\vert A)&=&\frac{1}{1-q}\left[1-\frac{{\rm Tr}(\rho^q(A,B))}{{\rm Tr}(\rho^q(A))}
\right]\nonumber \\
&=&\frac{1}{1-q}\left[1-\frac{\sum_n\lambda^q_n(A,B)}{\sum_m\,\lambda^q_m(A)}\right]
\end{eqnarray}
\noindent (where $\lambda_n(A,B),\  \lambda_m(A)$ are the eigenvalues of $\rho(A,B)$ and $\rho(A)$ 
respectively~\cite{note0}) 
and employed it to investigate the issue of 
separability. 
Tsallis $q$-conditional entropy method (AR approach) has also been  
employed to investigate separability in several finite dimesional quantum systems~\cite{Tsallis}.  As any 
spectral criteria, based only on the eigenvalues of the state and
its subsystems, do not provide a complete characterization of separability~\cite{vwolf}, the AR $q$-conditional 
entropy characterization does not lead, in general, to the necessary and 
sufficient criteria for separability. However, this  approach is 
fruitful in obtaining stronger criteria 
than the one derived from the familiar $q=1$ case~\cite{note0} i.e, the result based on von Neumann conditional 
entropy.

The AR $q$-entropy approach relies on finding the global and local spectra of the density matrices, which are 
not  straightforward  in the case of  CV systems. However, for $n$-mode Gaussian states, one 
can evaluate finite number ($n$) of symplectic eigenvalues~\cite{NM} of the 
corresponding $2n\times 2n$ variance matrix (which completely characterizes the Gaussian state) -- in terms of 
which the eigenvalues of the density matrix may be expressed readily~\cite{sirs, FI2, Lloyd}.  However,  
the issue of separability based on conditional $q$-entropy approach  has not been addressed so far in the 
context of Gaussian states, to the best of our knowledge. The present Brief Report aims towards investigating 
separability in Gaussian states based on AR $q$-entropic approch, thus filling an important gap.  


We consider $n$-mode Gaussian states, which are completely determined by the $2n\times 2n$ covariance matrix  
$V_{\alpha\beta}=\frac{1}{2}\, \langle \{\bigtriangleup\xi_{\alpha},\bigtriangleup\xi_{\beta}\}\rangle,\ 
\ \alpha,\beta=1,2,\ldots , 2n$;\  $\bigtriangleup\xi=\xi-\langle\xi\rangle$, $\{O_1, O_2\}= O_1\, O_2+ 
O_2\, O_1$  and $\langle  O\rangle~=~{\rm Tr}[\rho  O]$ denotes the expectation value of the operator $ O$. 
Under a $2n\times 2n$ symplectic transformation~\cite{NM} $S\in {\rm Sp}(2n, {\rm R})$, a Gaussian state is 
mapped to another Gaussian state characterized by the covariance matrix $V'=SVS^T.$ Then, it follows from  
Williamson theorem that for every covariance matrix $V$ there exists a symplectic matrix $S$ such that  
$SVS^T={\rm diag}(\nu_1,\nu_1; \nu_2,\nu_2; \ldots ; \nu_n,\nu_n),$
where $\nu_k, k=0,1,\ldots, n$ denote the symplectic eigenvalues~\cite{NM}.   
Correspondingly, the associated density matrix is expressed as a tensor product of $n$ thermal states 
of oscillators: 
\begin{eqnarray}
\label{sprho}
\rho_n&& \rightarrow \rho'_n=U(S)\, \rho_n \, U^\dag(S)=\bigotimes_{k=1}^n\, \rho(\nu_k) 
\end{eqnarray}
 where  
$\rho(\nu_k)=\frac{1}{\nu_k+\frac{1}{2}}\sum_{j=0}^\infty\left(\frac{\nu_k-\frac{1}{2}}{\nu_k+\frac{1}{2}}
\right)^j\, \vert j\rangle_k\langle j \vert.$  (Here $\{\vert j\rangle_k,\ j=0,1,\ldots, \infty\}$ denote the 
number states of the $k$th mode). An arbitrary positive power ${\rm Tr}[\rho^q],\ 0~<~q~\leq~\infty$ of 
the $n$-mode Gaussian density operator may thus be readily expressed in terms of the symplectic eigenvalues 
as~\cite{Lloyd}, 
\begin{eqnarray*}
{\rm Tr}[\rho^q_n] 
&=& \prod_{k=1}^n {\rm Tr}\left[\rho^q(\nu_k)\right]
=\prod_{k=1}^n  \frac{1}{(\nu_k+\frac{1}{2})^q-(\nu_k-\frac{1}{2})^q}. 
\end{eqnarray*}  
 
Considering a bipartite division  of a $n$~mode Gaussian system 
$\rho_n(A,B)$,  with marginals ${\rm Tr}_B[\rho_n(A,B)]=\rho_N(A),$ ${\rm Tr}_A[\rho_n(A,B)]=\rho_{(n-N)}(B)$ 
(where $A\rightarrow N\, {\rm modes},\ B\rightarrow (n-N)\ {\rm modes},\,  N<n$), 
the AR $q$-conditional entropy Eq.~(\ref{sba}) associated with Gaussian states is readily expressible 
in terms of  respective symplectic eigenvalues 
$\nu^{(AB)}_k$, $\nu^{(A)}_l$  of $\rho_n(A,B)$ and $\rho_N(A)$ as 
\begin{tiny}
\begin{eqnarray}
\label{arqg}
S_{q}(B\vert A)=\frac{1}{q-1}\left[1-
\frac{\displaystyle\prod_{l=1}^N\left[\left(\nu^{(A)}_l+\frac{1}{2}\right)^q-\left(\nu^{(A)}_l-\frac{1}{2}
\right)^q\right]}
{\displaystyle\prod_{k=1}^n\left[\left(\nu^{(AB)}_k
+\frac{1}{2}\right)^q-\left(\nu^{(AB)}_k-\frac{1}{2}\right)^q\right]}\right]
\end{eqnarray} 
\end{tiny}
The $q$-conditional entropy is necessarily positive, when the modes $A,B$ are separable. Negative values of 
$S_{q}(B\vert A)$ therefore imply entanglement between the modes A and B -- offering a sufficient condition to 
characterize entanglement in Gaussian states~\cite{RS,Duan}.  

On the other hand the PPT criterion translates itself to the following constraint:  the lowest symplectic 
eigenvalue $\tilde{\nu}_{\rm min}$ of the variance matrix $\tilde{V}$ (where  the canonical momenta $p_l$ of 
the  transposed  modes reverse their sign~\cite{RS}) of the partially transposed density matrix 
$\rho^T$ satisfies $\tilde{\nu}_{\rm min}\geq\frac{1}{2}$ for all separable Gaussian states~\cite{RS,RFW2}. 
Violation of this condition viz.,        
$\tilde{\nu}_{\rm min}< \frac{1}{2}$
is a characteristic of  entanglement. This PPT based characterization serves as a necessary and sufficient 
condition for separability in  two mode Gaussian states. 

To examine the utility of the AR $q$-entropy approach, we will discuss separability of mixed two-mode Gaussian 
states of physical importance. We compare the inseparability range obtained using the $q$-entropy criteria with 
that obtained using conditional von-Neumann entropy and also that resulting from PPT. 

\noindent {\em Two mode squeezed thermal state:} Density matrix of the two mode squeezed thermal state is given 
by~\cite{xyc}
\[
\rho(A,B)=U(S_r)\, \rho_{\rm th}(A)\otimes\rho_{\rm th}(B)\, U^{\dagger}(S_r).
\]
Here  $U(S_r)=\exp\left[\frac{r}{2}(a_1^\dagger a_2^\dagger-a_1a_2)\right]$  corresponds to the two-mode 
squeezing operator~\cite{plk}; $r$ is the real positive squeezing parameter  
and  $\rho_{\rm th}(A)$, $\rho_{\rm th}(B)$ denote single mode thermal states, both at same temperature $T$. 

The variance matrix $V(A,B)$ of the two-mode squeezed thermal state is given explicitly by 

{\scriptsize
\be 
V(A,B)=\frac{\coth(\beta/2)}{2}\left(\begin{array}{cccc} \cosh r & 0 & \sinh r & 0 \\ 
0 & \cosh r & 0 & -\sinh r \\ 
\sinh r & 0 &\cosh r & 0 \\
0 & -\sinh r & 0 & \cosh r\end{array}\right)
\ee}

\noindent where $\beta=T^{-1}$ is the inverse temperature (which is a dimensionless parameter with the choice of 
appropriate 
units i.e., $\hbar$, the oscillator frequency $\omega$ and the Boltzmann constant $\kappa$ are equal to one).  
The symplectic eigenvalues $\nu_k^{(AB)},\ k=1,2$ associated with this state 
are degenerate and are given by $\nu_{k=1,2}^{(AB)}=\frac{\coth(\beta/2)}{2}$.  
The symplectic eigenvalue of the reduced density matrix $\rho(A)$ is found to be  $\nu(A)=\frac{\coth(\beta/2)\, 
\cosh r}{2}$.
Now, using Eq.~(\ref{arqg}), the conditional q-entropy associated with two mode squeezed thermal state 
may be readily 
obtained as
\begin{tiny} 
\begin{eqnarray}
\label{sba1}
S_{q}(B\vert A)=\frac{1}{q-1}\left[ 
\frac{\left(\coth(\beta/2)\, \cosh r+1\right)^q-\left(\coth(\beta/2)\, \cosh r-1\right)^q}
{\left[\left(\coth(\beta/2)+1\right)^q-\left(\coth(\beta/2) -1\right)^q\right]^2}\right]    
\end{eqnarray}
\end{tiny} 

\begin{figure}[h]
\label{3}
\includegraphics*[width=2.2in,keepaspectratio]{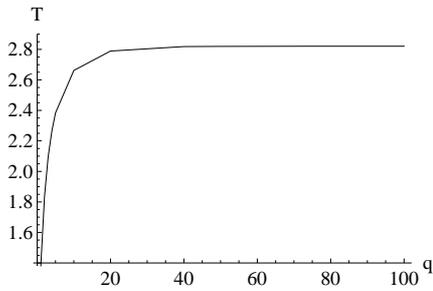}
\caption{ Implicit plot of $S_q(B\vert A)=0$ (with the choice of the parameter $ r=2$) as function of $q$ for 
the two-mode squeezed thermal state.}
\end{figure}
An implicit plot of $S_q(B\vert A)=0$ (see 
Fig.~1) shows that $T_c^{(\infty)}\rightarrow 2.82$ in the limit  
$q\rightarrow \infty$, for $r=2$. One can also see that the 
temperature $T^{(1)}_c\approx 1.381$ above which the conditional von-Neumann entropy $S_{1}(B\vert A)$ is 
positive. It is clear that  the threshold temperature $T_c$ increases with increasing $q$ and the strongest 
limitation on separability results when $q\rightarrow \infty$. 

In order to compare the effectiveness of the AR q-entropic characterization with that based on the PPT 
criterion, we identify that the minimum symplectic eigenvalue of  the partially transposed squeezed thermal 
state   $\tilde\nu_{\rm min}~=~\frac{1}{2}\,e^{-r}\coth{\frac{\beta}{2}}$ is less than $\frac{1}{2}$  when 
 $T^{\rm PPT}_c \geq 3.672$,  for  $ r=2$.  
This clearly reveals that the separability domains inferred via the threshold temparature values follow the 
trend $T^{(1)}_c<T^{(\infty)}_c<T^{\rm PPT}_c$. In other words, the PPT criterion 
gives the strongest limitation~\cite{pptlimit} (which is both necessary and sufficient) on separability.

\noindent {\em Two-mode state resulted by combining  a squeezed state 
and a thermal state in a 50:50 beam splitter:} Now we consider a two mode Gaussian state obtained  when a single 
mode squeezed state  interferes with a single mode thermal state through a $50:50$ beam splitter~\cite{FI2}.  
The variance matrix of the resulting two-mode state is given by~\cite{FI2}   
\be
V(A,B)=\frac{1}{4}\ba{cccc} a+b & 0 & a-b & 0 \\ 
                                0 & a+\frac{1}{b} & 0 &  a-\frac{1}{b} \\
                                a-b & 0 & a+b & 0 \\
0 & a-\frac{1}{b} & 0 &  a+\frac{1}{b} \ea
\ee
Here $b=e^\eta$ with $\eta$ denoting the single mode squeezing parameter  and $a=\coth(\beta/2)$,  where 
$\beta=T^{-1}$ corresponds to the inverse temperature of the input thermal state. 

The symplectic eigenvalues of $V(A,B)$ are non-degenerate and are found to be 
\be
\label{vab2}
\nu^{(AB)}_1=\frac{1}{2};\ \   \nu^{(AB)}_2=\frac{1}{2}\coth{\frac{\beta}{2}}.
\ee 
The symplectic eigenvalue of $V(A)$ (and also $V(B)$) is found to be 
\begin{tiny}
\be
\label{va2}
\nu^{(A)}=\frac{1}{4}\sqrt{(a+b)(a+\frac{1}{b})}=\frac{1}{4}\sqrt{1 + 2 \cosh{\eta} 
\coth{\frac{\beta}{2}}+\coth^2{\frac{\beta}{2}}}. 
\ee
\end{tiny}
Substituting Eqs.~(\ref{vab2}) and (\ref{va2}) in (\ref{arqg}) one can obtain an explicit expression for 
$S_q(B\vert A)$.
\begin{figure}
\label{5}
\includegraphics*[width=2.2in,keepaspectratio]{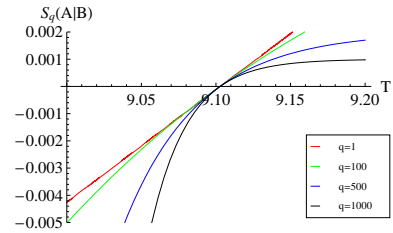}
\caption{ The conditional q-entropy  $S_q(B\vert A)$ for different values of $q$, 
as a function of temperature $T~=~\beta^{-1}$ 
of the Gaussian state resulting by combining a single mode squeezed state (with squeezing parameter $\eta=4$) 
with a  thermal state in a $50:50$ 
beam splitter.}
\end{figure}

In Fig.~2 we illustrate the variation of $S_q(B\vert A)$ for different choices of the parameter $q$.     
It is evident from Fig.~2 that the AR q-entropy ceases to be negative at $T_c \approx 9.1$ for different 
choices of $q$. In other words, both conditional von-Neumann entropy $S_1(B\vert A)$  and the AR $q$ 
entropy,  in the limit $q\rightarrow \infty$, lead to the same inseparability range. The exact  
inseparability range obtained by  PPT criterion is much stronger (for the choices of the parameters we find that 
the condition $\tilde\nu_{\rm min}=1/2$ on the lowest eigenvalue of $\tilde{V}(A,B)$ is satisfied  for 
$T^{\rm PPT}_c \approx 27.3$). Thus we find that  $T^{(1)}_c=T^{(\infty)}_c < T^{\rm PPT}_c$ for the state under 
consideration. 

\noindent{\em Two mode Squeezed state subjected to a coupled leaky wave guide:} As our third example, we 
consider a mixed two-mode Gaussian state that results when a pure 
two-mode squeezed state is transmitted via 
a coupled waveguide system with non-zero leakage~\cite{GSA}. 

Transmission of a pure two mode squeezed vacuum state $\exp\left[\frac{r}{2}\, (a_1^\dag a_2^\dag-a_1\, 
a_2)\right]\vert 0, 
0\rangle$ via two leaky waveguides coupled to each other by an interaction term $H_{\rm int}~=~
{\cal J}\, (a_1^\dag a_2+a_2^\dag\, a_1),$ (${\cal J}$ denotes the coupling strength), results in a mixed two 
mode Gaussian state, the variance matrix $V(A,B)$ of which is given by~\cite{GSA} 
\be
\label{v3}
V(A,B)=\ba{cccc} f & g & h & 0 \\ 
                                g & f & 0 &  -h \\
                                h & 0 & f & g \\
0 & -h & g &  f \ea
\ee
where 
$f$=$\frac{1}{2}+e^{-2\gamma t}\sinh^2(\frac{ r}{2})$, $g$=-$\frac{1}{2}e^{-2\gamma t} \sinh {r}\sin(2{\cal J}t)$ and  
$h=\frac{1}{2}e^{-2\gamma t} \sinh {r}\cos(2{\cal J}t).$

Here $ r$ denotes the squeezing parameter of the input state and  $\gamma$ corresponds to 
leakage (decay rate) of individual  modes. Whereas the global symplectic spectra of $V(A,B)$ are given by 
$\nu^{(AB)}_{1}=\nu^{(AB)}_{2}=\sqrt{f^2-g^2-h^2}$, its local symplectic eigenvalue is 
$\nu^{(A)}=\sqrt{f^2-g^2}$.
\begin{figure}[h]
\label{7}
\includegraphics*[width=2.2in,keepaspectratio]{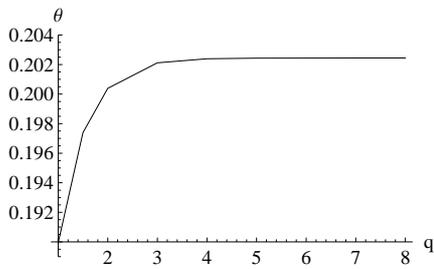}
\caption{An implicit plot of $S_q(B\vert A)=0$ as a function of $q$ for the two-mode squeezed state subjected to 
a coupled leaky wave guide. (Here, $r=1.8$ and $\gamma/{\cal J}=0.1$).}
\end{figure}
In fact, the time evolution of entanglement of the two-mode Gaussian state transmitted via 
 a coupled leaky wave guide exhibits  a damped oscillation pattern~\cite{GSA}, and this oscillatory behavior 
repeats until a total decay takes place due to environmental decoherence. 
In the present discussion we have focussed only on the time duration for which the initially 
entangled two mode squeezed state loses its entanglement at a first glance during evolution. We find that 
the scaled time $\theta={\cal J}t/\pi$ for which the conditional von-Neumann entropy $S_1(B\vert A)$ ceases to be negative is 
given by $\theta_c^{(1)}\approx 0.19$, when $r=1.8$ and $\gamma/{\cal J}=0.1$. Stronger 
limitations on separability follow with the increase of the parameter $q$.  It is evident  from Fig.~3 that 
the largest scaled time interval $\theta_c^{(q)}$ (which approaches the value 0.202), after which the initially 
entangled two mode  state becomes separable,  is realized in the limit $q\rightarrow \infty.$   
On the other hand the necessary and sufficient condition for separability (identified by the condition 
$\tilde\nu_{\rm min}\geq 1/2$ on   
the smallest symplectic eigenvalue $\tilde\nu_{\rm min}$) leads to threshold scaled time $\theta_c^{\rm 
PPT}\approx 0.23$.    
Thus, it follows that $\theta_c^{(1)}<\theta_c^{(\infty)}<\theta_c^{\rm PPT}.$ 

In conclusion, we have explored separability in 
Gaussian states based on the AR $q$-conditional entropy approch. 
This is facilitated by expressing the $q$-conditional entropy 
in terms of the  symplectic eigenvalues of the state. We have analyzed the separability features of three 
different examples of  two-mode Gaussian states using this entropic approach and compared the results  with 
those obtained from conditional von-Neumann entropy ($q=1$ limit of AR $q$-entropy) and with the PPT method. 
Strongest limitation 
on separability is realized in the limit $q\rightarrow \infty$, although the $q$-entropy approach leads to 
weaker domain of separability than the exact one obtained from PPT method.       


\begin{thebibliography}{1}
\bibitem{note0} Note that in the limit $q\rightarrow 1$ the AR $q$-conditional entropy Eq.~(\ref{sba}) reduces 
to von Neumann conditional entropy i.e.,  $\displaystyle\lim_{q\rightarrow 1}S_q(B\vert A)\rightarrow S(B\vert 
A)=S(A,B)-S(B)$.
\bibitem{pptlimit} The values of the threshold temperatures 
$T^{(1)}_c$, $T^{(\infty)}_c$ and $T^{\rm PPT}_c$ depend  on the 
squeezing parameter $ r$ and it may be noted that   
for a higher value of squeezing parameter $ r=5$, the threshold temperatures take the values  
$T^{(1)}_c \approx 27.3$,  $T^{(\infty)}_c \approx 73.2$, while 
$T^{\rm PPT}_c\approx 74.2.$ In other words,  the separability range 
inferred by $q$-conditional entropy method (in the limit $q\rightarrow \infty$) approaches closer 
to the exact one based on PPT for larger values of the squeezing parameter.
\end{thebibliography}
\end{document}